\documentstyle[12pt]{article}

\advance\voffset by -1.5cm
\advance\hoffset by -1.8cm
\input amssym.def
\input amssym.tex
\def\be{\begin{equation}}
\def\ee{\end{equation}}

\def\Zop{{\Bbb Z}}

\def\pmb#1{\setbox0=\hbox{#1}%
 \kern-.025em\copy0\kern-\wd0
 \kern.05em\copy0\kern-\wd0
 \kern-.025em\raise.0433em\box0 }

\def\bz{{\bar{z}}}

\def\J{{\cal J}}
\def\bJ{{\bar{\cal J}}}
\def\B{{\cal B}}
\def\M{{\cal M}}

\def\H{{\cal H}}

\def\z{\zeta}
\def\bH{{\bar{H}}}

\def\d{{\partial}}
\def\bd{{\bar{\partial}}}

\def\bE{{\bar E}}

\def\t{\theta}

\def\ttheta{{\tilde{\theta}}}

\def\lg{{\frak g}}

\def\3{\ss}
\def\sq{\hbox{\rlap{$\sqcap$}$\sqcup$}}
\def\qed{\ifmmode\sq\else{\unskip\nobreak\hfil
\penalty50\hskip1em\null\nobreak\hfil\sq
\parfillskip=0pt\finalhyphendemerits=0\endgraf}\fi}
\def\half {\frac{1}{2}}

\def\bbbc{{\mathchoice {\setbox0=\hbox{$\displaystyle\rm C$}\hbox{\hbox
to0pt{\kern0.4\wd0\vrule height0.9\ht0\hss}\box0}}
{\setbox0=\hbox{$\textstyle\rm C$}\hbox{\hbox
to0pt{\kern0.4\wd0\vrule height0.9\ht0\hss}\box0}}
{\setbox0=\hbox{$\scriptstyle\rm C$}\hbox{\hbox
to0pt{\kern0.4\wd0\vrule height0.9\ht0\hss}\box0}}
{\setbox0=\hbox{$\scriptscriptstyle\rm C$}\hbox{\hbox
to0pt{\kern0.4\wd0\vrule height0.9\ht0\hss}\box0}}}}

\begin{document}
\thispagestyle{empty}
\begin{flushright}
DAMTP-95-72 \\
hep-th/9601016
\end{flushright}
\vspace{2.0cm}

\begin{center}

{\Large {\bf Abelian duality in WZW models}}
\vspace{2.0cm}

{\large Matthias R. Gaberdiel} 
\footnote{e-mail: M.R.Gaberdiel@damtp.cam.ac.uk} \\
{Department of Applied Mathematics and Theoretical
Physics\\
University of Cambridge, Silver Street \\
Cambridge, CB3 9EW, U.\ K.\ }
\vspace{0.5cm}

December 1995
\vspace{3.0cm}

{\bf Abstract}
\end{center}

{\leftskip=2.4truecm
\rightskip=2.4truecm

We analyse abelian $T$-duality for WZW models of simply-connected
groups. We demonstrate that the dual theory is a certain orbifold of
the original theory, and check that it is conformally invariant. We
determine the spectrum of the dual theory, and show that it agrees
with the spectrum of the original theory.

}

\newpage

\section{Introduction}
\renewcommand{\theequation}{1.\arabic{equation}}
\setcounter{equation}{0}

Recently there has been a renewed interest in duality transformations
in field theory and string theory \cite{SW,W2,S}. These transformations
relate the strong coupling region of one theory to the weak coupling
regime of another theory, the dual theory. The former is typically
not accessible by perturbation theory, whereas the latter might well
be. Duality therefore seems to provide a very promising tool
for the non-perturbative analysis of field theories and string
theories. 

The archetypal prototype of this transformation in string theory, is
what is now usually referred to as $T$-duality. This is the
transformation which exchanges the winding and momentum modes of a
free boson whose target space is a circle. The transformation
also maps the radius $R$ of the circle to its inverse, and thus
relates phenomena at large and small scales. 

In this paper we want to study a slightly more complicated variation
of this transformation in detail, a chiral abelian duality
transformation of a WZW model. In this context the r\^ole of the
target space circle is played by the Cartan torus of a simple
group. This torus is embedded in the group in a rather non-trivial
fashion, and, as a consequence, the analysis of the duality
transformation is much more involved. On the other hand, these models
are still sufficiently simple to allow a detailed analysis. In
particular, both the original and the dual theory possess infinite
dimensional symmetry algebras, and the theories are therefore
essentially determined by the representation theory of these
algebras. For example, it is rather easy to see directly that the dual
theory is conformally invariant, and the calculation of the spectrum
is a fairly tractable algebraic problem.

In the archetypal situation, it was clear how to define the duality
transformation from the structure of the spectrum of the theory. In
the situation at hand, this is not the case, and we therefore follow
the procedure of Buscher \cite{Buscher,RV} to derive an action for the
dual theory from the classical action. We then quantise this action,
following Witten \cite{Wit}. It turns out that the dual theory has two
commuting infinite dimensional symmetry algebras, a (right-moving)
untwisted Kac-Moody algebra, and a (left-moving) twisted Kac-Moody
algebra, whose (inner) twists we determine. As the twist is inner, the
twisted algebra is isomorphic to the untwisted algebra.

The classical action has typically more than one quantisation
\cite{MG5}, and we want to analyse whether there is one such
quantisation for which the spectrum of the dual theory reproduces the
original spectrum. This is a rather non-trivial problem, as the
spectrum of a twisted algebra typically differs from the spectrum of
the corresponding untwisted algebra, even if they are related by inner
twists and thus isomorphic as algebras. In the case at hand, however,
due to the rather special form of the twists, such a quantisation may
be found. The resulting dual theory (which is then equivalent to the
original theory) can be naturally interpreted as an orbifold of the
original theory.
\smallskip

It should be mentioned that these models have already been considered
in \cite{Kir,AABL} (for a review see also \cite{AAL3}), where,
however, the structure of the dual theory was not analysed in
detail. Furthermore, in \cite{Kir} the global structure of the
(target space) group was not properly taken into account, and as a
consequence, the duality transformation was misidentified with a Weyl
transformation (see also \cite{AAL2}).  
\smallskip

The paper is organised as follows. In section~2, we derive the action
for the dual theory, paying particular attention to the global
properties of the target space group. We then quantise this action in
section~3, and determine the twist of the left-moving KM algebra. In
section~4, we construct the quantisation for which the spectrum is
preserved, and section~5 contains some concluding remarks.

\section{The duality transformation}
\renewcommand{\theequation}{2.\arabic{equation}}
\setcounter{equation}{0}

Let us consider the WZW action \cite{Wit,FGK,MPHOS} 
\be
\label{action}
S_0 [g] = -\frac{k}{4 \pi} \int_{\M} dz d\bz \; 
Tr \left( g^{-1} \d g g^{-1} \bd g \right) 
+ \frac{k}{12 \pi} \int_{\B} d^3 y \; \varepsilon^{ijk} \; 
Tr \left( g^{-1} \partial_i g\; g^{-1} \partial_j g 
\; g^{-1} \partial_k g \right) \,, 
\ee 
where $g: \M \rightarrow G$, $\M$ is a closed two-dimensional
manifold, and $\B$ is a three-dimensional manifold with boundary
$\M$. The trace is the Killing form on the algebra $\lg$ of the Lie
group $G$, normalised so that the long roots have length square equal
to $2$. We shall only consider the case where $G$ is simply connected;
then the quantisation condition on $k$ is that it is an integer. We
want to parametrise $g_0$ as 
\be
\label{para}
g_0 = e^{i \theta^\alpha e_\alpha} g \,,
\ee
where the sum extends from $\alpha=1$ to $n$, and the $e_\alpha$ are a
basis for a ($n$-dimensional) subalgebra of a Cartan subalgebra. Here
$n\leq r$, the rank of the Lie group $G$, the $\t^\alpha$ are
allowed to vary independently, and $g$ lies in some coset of $G$. We
could also consider the case, where the Cartan torus is 
multiplied from the right, and this would lead to a completely
analogous discussion. 
\smallskip

The manifold $\M$ need not be simply-connected and can in fact be of
any genus. Here we shall mainly consider the case where $\M$ is a torus,
as this is sufficient for the determination of the spectrum.
Typically, the fields $\t^\alpha$ have then non-trivial winding, and we 
denote by $\Delta^c \t^\alpha$ the monodromy of $\t^\alpha$ along the
cycle $c$, where $c=a$ or $c=b$, the two non-trivial cycles of the
torus. The consistency condition is then 
\be
\label{origper}
\frac{1}{2 \pi } \sum_{\alpha=1}^n \Delta^c \t^\alpha e_\alpha \in R^* \,,
\ee
where $R^*$ is the coroot lattice of the Lie algebra. (If $G$ were not
simply-connected, the left-hand side of (\ref{origper}) would only
have to be in the coweight lattice.) As the $\t^\alpha$ are independent
variables, this should imply that for each $\t^\alpha$ there exists a
minimal monodromy $\Delta^{m} \t^\alpha$ such that 
\be
\label{origper1}
\frac{1}{2 \pi} \Delta^{m} \t^\alpha e_\alpha \in R^* 
\ee
is a generating element of the coroot lattice ({\it i.e.}\ no shorter
multiple of it is in the coroot lattice), and all other monodromies
are integer multiples of the minimal one. To simplify notation, we
shall thus assume that the $e_\alpha$ are roots of the Lie algebra
$\lg$. Then (\ref{origper1}) is equivalent to
\be
\label{origper2}
\Delta^{m} \t^\alpha =  2 \pi  \frac{2}{Tr(e_\alpha e_\alpha)} \,.
\ee

We want to use the Polyakov-Wiegmann property (PW) \cite{PW}  
\be
\label{PW}
S_0[g_1 g_2] = S_0[g_1] + S_0[g_2] - \frac{k}{2 \pi} \int_{\M} dz d\bz\; 
Tr \left( g_1^{-1} \bd g_1 \; \d g_2 g_2^{-1} \right) \,,
\ee
to rewrite the action as 
\be
\label{ori}
S_0[g_0] = S_0[g] + {k \over 4 \pi} \int_{\M} dz d\bz \;
d_{\alpha \beta} \d \theta^\alpha \bd \theta^\beta 
- {k \over 4 \pi} \int_\M dz d\bz \;  J_\alpha 
\bd \theta^\alpha \,, 
\ee 
where
$$
d_{\alpha \beta} = Tr (e_\alpha e_\beta) \,, \hspace*{1.5cm} \mbox{and}
\hspace*{1.5cm} J_\alpha = 2 i \; Tr (\d g g^{-1} e_\alpha) \,.
$$

It should be noted that this simple form of the (PW) property only
holds if $\M$ is simply-connected\footnote{The remark in 
\cite[p.\ 81]{AABL} is not quite correct.}; in general it has to be
modified by the inclusion of a topological term involving $g_1$ and
$g_2$ \cite[Appendix 2, eq.\ (10)]{FGK}. For example, we can calculate
the action for $g_1\, g_2$, where $g_j=\exp(i\t^j e_j)$ is in the
Cartan torus, in two different ways, directly from the action, and
using the (PW) formula   
\be
S_0^{(PW)}[g_1 g_2] - S_0[g_1 g_2]  = \frac{k}{4 \pi} \int_\M dz d\bz \;
Tr(e_1 e_2) \left( \bd \t^1 \d \t^2 - \bd \t^2 \d \t^1 \right) \,.
\ee
The difference is a topological term which vanishes for
simply-connected world-sheets. If the world-sheet has the topology of
a torus, however, we can integrate by parts to obtain
\be
S_0^{(PW)}[g_1 g_2] - S_0[g_1 g_2] = 
\frac{Tr(e_1 e_2) k }{4 \pi} \left(  \Delta^a \t^1 \Delta^b \t^2
- \Delta^b \t^1 \Delta^a \t^2 \right) \,.
\ee
(\ref{origper1}) now implies that this is in $\pi k \Zop$, as the
Killing form of two elements in the coroot lattice is integral, but it
is not always in $2 \pi \Zop$. On the other hand, it should be clear
that a topological term involving $e^{i \theta^\alpha e_\alpha}$ and
$g$ will not affect the following argument. (Such a term will only
restrict the allowed monodromies for $g$ --- {\it cf.}\ the
discussion after (\ref{partint}).)
\medskip

Originally, the duality transformation was formulated in terms of
gauging the action and fixing the gauge by introducing Lagrange
multipliers. Upon integrating out the Lagrange multipliers, the gauge
fields are restricted to be flat, and we can choose the monodromies of
the Lagrange multipliers, so that the monodromies of the gauge fields
lie in a lattice. For compact gauge groups we can thus guarantee that
the gauge fields are trivial, and the original action can be
recovered. On the other hand, integrating out the gauge fields leads to
a new action in which the Lagrange multipliers take the r\^ole of the
fields whose symmetry was gauged \cite{Buscher,AABL,RV}.

It was realised recently \cite{AAL2} that the duality
transformation can also (formally) be obtained as a certain canonical
transformation. In this approach, it is not necessary to gauge the
action, but the transformation with respect to $\theta^1$, say, is 
simply the transformation 
\be
\label{canon}
\int D \theta^1 \;\; \exp\left( {\lambda k i \over 4 \pi} 
\int_\M dz d\bz \; (\bd \theta^1 \d \psi^1 
- \d \theta^1 \bd \psi^1)\right)  \; \exp\left(i S_0[g_0] \right) \,,
\ee
where $\lambda$ is an arbitrary constant which parametrises the
freedom to rescale $\psi^1$. To calculate this transformation, we
first rewrite the action as  
\begin{eqnarray}
S_0[g_0] & = & S_0[g] + \frac{k}{4 \pi} \int_\M dz d\bz \; \left[
d_{11} \d \t^1 \bd \t^1 - J_1 \bd \t^1
+ \sum_{\alpha\geq 2} \left( d_{1\alpha} \d \t^1 \bd \t^\alpha
+ d_{\alpha 1} \bd \t^1 \d \t^\alpha  \right) \right] \nonumber \\
& & \;\;\;\;
+ \frac{k}{4 \pi} \int_\M dz d\bz \left[
\sum_{\alpha,\beta\geq 2} d_{\alpha \beta} \d \t^\alpha \bd \t^\beta 
- \sum_{\alpha\geq 2} J_\alpha \bd \t^\alpha \right]\,.
\label{orii}
\end{eqnarray}
We then add the generating functional for the canonical transformation
and complete the square to obtain
$$
S_0[g_0] + {\lambda k \over 4 \pi} \int_\M dz d\bz \;
(\bd \theta^1 \d \psi^1 - \d \theta^1 \bd \psi^1) \hspace*{9.5cm} 
\vspace*{-0.5cm}
$$
\begin{eqnarray}
& = & S_0[g] + {k \over 4\pi} \int_\M dz d\bz \; d_{11} 
\left( \d \theta^1 + \frac{d_{\alpha 1}}{d_{11}} \d \t^\alpha 
-  \frac{J_1}{d_{11}}  + \frac{\lambda}{d_{11}} \d \psi^1\right) 
\left(\bd \theta^1 + \frac{d_{1\alpha}}{d_{11}} \bd \t^\alpha
- \frac{\lambda}{d_{11}} \bd \psi^1 \right) \nonumber \\
& & \;\;\; + {k \over 4\pi} \int_\M dz d\bz \left[ 
{\lambda^2 \over d_{11}} \d \psi^1 \bd \psi^1 
- {\lambda \over d_{11}} J_1 \bd \psi^1 \right] \nonumber \\
& & \;\;\; + {k \over 4\pi} \int_\M dz d\bz \left[ 
{\lambda d_{\alpha 1} \over d_{11}} \d \theta^\alpha \bd \psi^1 
- {\lambda d_{1\alpha} \over d_{11}} \d \psi^1 \bd \theta^\alpha \right]
\nonumber \\
& & \;\;\; + {k \over 4\pi} \int_\M dz d\bz \left[ 
\left( d_{\alpha \beta} - {d_{\alpha 1} d_{1\beta} \over
d_{11}}\right) \d \theta^\alpha \bd \theta^\beta 
- \left( J_\alpha - {d_{1 \alpha} \over d_{11}} J_1 
\right) \bd \theta^\alpha \right] \,,
\end{eqnarray}
where a summation over $\alpha$ and $\beta$ from $2$ to $n$ is
implicit. We shift the variable of integration $\theta^1$, without
incurring a Jacobian factor, to obtain a purely quadratic term in
$\theta^1$. Upon integration this gives only a constant, and the dual
action is therefore
\begin{eqnarray}
\label{dualtheo}
S^\natural & = & S_0[g] + {k \over 4\pi} \int_\M dz d\bz \left[ 
{\lambda^2 \over d_{11}} \d \psi^1 \bd \psi^1 
- {\lambda \over d_{11}} J_1 \bd \psi^1 \right] \nonumber \\
& & \;\;\; + {k \over 4\pi} \int_\M dz d\bz \left[ 
{\lambda d_{\alpha 1} \over d_{11}} \d \theta^\alpha \bd \psi^1 
- {\lambda d_{1\alpha} \over d_{11}} \d \psi^1 \bd \theta^\alpha \right]
\nonumber \\
& & \;\;\; + {k \over 4\pi} \int_\M dz d\bz \left[ 
\left( d_{\alpha \beta} - {d_{\alpha 1} d_{1\beta} \over
d_{11}}\right) \d \theta^\alpha \bd \theta^\beta 
- \left( J_\alpha - {d_{1 \alpha} \over d_{11}} J_1 
\right) \bd \theta^\alpha \right] \,.
\end{eqnarray}
It is believed that the dual theory is conformally invariant, provided
that the dilaton is suitably transformed \cite{Buscher}. For the
example we are considering here, we shall also see this more
explicitly later.

If we integrate (\ref{canon}) by parts, only the
boundary term contributes which is, on the torus, of the form
\be 
\label{boundary}
\frac{\lambda k i}{4 \pi}
 \left( \Delta^a \psi^1 \Delta^b \t^1
- \Delta^b \psi^1 \Delta^a \t^1 \right) \,.
\ee
In the path integral we integrate over all $\t^1$ satisfying
(\ref{origper1}) for $\alpha=1$. The group $G$ is simply-connected, 
and the different components in the integral, parametrised by the different
(allowed) values for $\Delta^c \t^1$, should therefore all give the same
contribution. As only $\exp(i S)$ is relevant in the path integral,
this implies that the dual action is only non-zero for $\psi^1$
satisfying ({\it cf.}\ Alvarez {\it et.al.}\ \cite{AAL2})
\be
\label{dualboun}
\frac{\lambda k }{4 \pi} \Delta^d \psi^1 \Delta^{m} \t^1 \in 2 \pi
\Zop \,.  
\ee

Before we start interpreting this dual theory, let us note that
the duality transformation can be reversed by the inverse
transformation 
$$
\int D \psi^1 \;\; \exp\left( {\lambda k i \over 4 \pi} \int_\M dz d\bz \;
(\bd \psi^1 \d \ttheta^1 - \d \psi^1 \bd \ttheta^1)\right)  
\exp(i S^\natural)\,.
$$
Indeed, completing the square in the sum of the dual action and the
generating functional for the second transformation, we find 
$$
S^\natural + {\lambda k \over 4 \pi} \int_\M dz d\bz \;
(\bd \psi^1 \d \ttheta^1 - \d \psi^1 \bd \ttheta^1)  \hspace*{10.5cm} 
\vspace*{-0.6cm}
$$
\begin{eqnarray}
& = & S_0[g] + {k \over 4\pi} \int_\M dz d\bz \; 
{\lambda^2 \over d_{11}}
\left( \d \psi^1 + \frac{d_{\alpha 1}}{\lambda} \d \t^\alpha
-  \frac{J_1}{\lambda}
+ \frac{d_{11}}{\lambda} \d \ttheta^1\right) 
\left(\bd \psi^1 - \frac{d_{1\alpha}}{a} \bd \t^\alpha
- \frac{d_{11}}{\lambda} \bd \ttheta^1 \right) \nonumber \\
& & \;\;\; + {k \over 4\pi} \int_\M dz d\bz \left[ 
d_{11} \d \ttheta^1 \bd \ttheta^1 
- J_1 \bd \ttheta^1 \right] 
+ {k \over 4\pi} \int_\M dz d\bz  \;
\left( d_{1\alpha} \d \ttheta^1 \bd \t^\alpha
+ d_{\alpha 1} \d \t^\alpha \bd \ttheta^1 \right)  \\
& & \;\;\; + {k \over 4\pi} \int_\M dz d\bz \left[ 
\left( d_{\alpha \beta} - {d_{\alpha 1} d_{1\beta} \over
d_{11}} + {d_{\alpha 1} d_{1\beta} \over d_{11}}
\right) \d \theta^\alpha \bd \theta^\beta 
- \left( J_\alpha - {d_{1 \alpha } \over d_{11}} J_1 
+ {d_{1 \alpha} \over d_{11}} J_1 \right) \bd \theta^\alpha \right] \,,
\nonumber
\end{eqnarray}
which equals the original action. Furthermore, we also have  
\be
\frac{\lambda k }{4 \pi} \Delta^{m} \psi^1 \Delta^d \ttheta^1 \in 2
\pi \Zop \,,
\ee
and thus the periodicity of $\ttheta^1$ is the same as that of $\t^1$,
because of (\ref{dualboun}).
\medskip

The term in the second line of (\ref{dualtheo}) is topological and 
vanishes on world-sheets with trivial topology. Apart from this
term we can identify the dual action (\ref{dualtheo}) with the
original action (\ref{ori}). Indeed, if we choose (without loss of
generality) $\lambda=d_{11}$, and parametrise the group as   
\be 
g_0 = e^{i \psi^1 \hat{e}_1} \left( \prod_{\alpha=2}^{n} 
e^{i \t^\alpha \hat{e}_\alpha} \right) g \,, 
\ee 
where
\be
\label{transf}
\hat{e}_\alpha := e_\alpha - { Tr (e_\alpha e_1) \over Tr (e_1 e_1)}
\;  e_1  \hspace*{0.5cm} \mbox{for $\alpha\geq 2$,} \hspace*{2cm}
\hat{e}_1 := e_1 \,,
\ee
then the two actions coincide. However, the minimal
monodromy of the field $\psi^1$ (for this choice of $\lambda$)   
\be
\label{chimon}
\Delta^m \psi^1 = 4 \pi \frac{d_{11}}{2 \lambda k} = \frac{2 \pi}{k}
\ee 
is in general different from the minimal monodromy of $\t^1$ before
(\ref{origper2}). (The monodromy is only the same if $\t^1$ is a long
root, and $k=1$\footnote{In this case there exists a free field
construction for which $e_1$ is represented by a free boson
\cite{FKS,GNOS}. At least in the case of $\lg= \frak s \frak u (2)$,
this free field theory is self-dual with respect to the ordinary
duality transformation of a free boson on a circle \cite{Ginsparg}.}.) 
In general, the dual action therefore does not simply correspond to a
redefinition of the original fields (or a Weyl transformation as
advocated by Kiritsis \cite{Kir}), but rather to a certain non-trivial
orbifold \cite{DHVW}. 

The above topological term guarantees that the monodromy of the other
fields is in the coroot lattice up to integral multiples of
(\ref{chimon})\footnote{This is automatically true if $k$ is a 
multiple of $d_{11}$, as $\Delta^c\t^\alpha\,\hat{e}_\alpha$ is always
in the coroot lattice up to $k/d_{11} \; \Delta^m \psi^1 \,e_1$.}. To
see this we write $\Delta^c \psi^1 =n^c \Delta^m \psi^1$ with $c=a,b$
and $n^a,n^b\in\Zop$; the topological term then becomes 
\be
\label{partint}
\frac{1}{2} \sum_{\alpha=2}^{n} \left( n^a 
Tr(\Delta^b \t^\alpha e_\alpha \;e_1) -  n^b 
Tr(\Delta^a    \t^\alpha e_\alpha \;e_1) \right) \,.
\ee
As we calculate the partition function of the theory, {\it i.e.}\ the
path-integral of $\exp(iS^\natural)$ over the torus, we have to sum
over all $n^a, n^b\in\Zop$. Then (\ref{partint}) will imply that
only those states contribute, whose monodromy
$\Delta^c\t^\alpha\,\hat{e}_\alpha$ satisfies the above
condition. This then means that the dual action describes the
orbifold of the original theory which is induced by the subgroup
generated by         
\be 
\exp\left(\frac{2 \pi i} {k} e^1 \right)\,.  
\ee
\smallskip

We should also mention that the precise from for the redefinition of
the fields (\ref{transf}) is a matter of choice: we can integrate the
original action (\ref{orii}) by parts to rewrite it as 
\begin{eqnarray}
S_0[g_0] & = & S_0[g] + \frac{k}{4 \pi} \int_\M dz d\bz \left[
d_{11} \d \t^1 \bd \t^1 - J_1 \bd \t^1 \right] \nonumber \\
& & \;\;\;\; + \frac{k}{4\pi} \int_\M dz d\bz \left[
+ \sum_{\alpha \geq 2} \left( (1-m_\alpha) d_{1 \alpha} \d \t^1 \bd
\t^\alpha + (1+m_\alpha) d_{\alpha 1} \bd \t^1 \d \t^\alpha \right)
\right] \nonumber \\ 
& & \;\;\;\;
+ \frac{k}{4 \pi} \int_\M dz d\bz \left[
\sum_{\alpha,\beta\geq 2} d_{\alpha \beta} \d \t^\alpha \bd \t^\beta 
- \sum_{\alpha\geq 2} J_\alpha \bd \t^\alpha \right]\,,
\label{oriii}
\end{eqnarray}
where $m_\alpha\in\Zop$ and we have ignored the topological terms. We
can then do the same analysis as before and find that, apart from the
new monodromy property (\ref{chimon}) for the field $\psi^1$ and some
topological terms, the dual action agrees with the original action,
where we have parametrised 
\be
\label{transfi}
\hat{e}_\alpha := e_\alpha - {(1+m_\alpha) Tr(e_1 e_\alpha) \over 
Tr (e_1 e_1)} e_1  \hspace*{0.5cm} \mbox{for $\alpha\geq 2$,} 
\hspace*{2cm} \hat{e}_1 :=  e_1 \,.
\ee
Again, this indicates that the characteristic feature of the duality
transformation is the modification of the monodromy properties, rather
than a redefinition of the fields.

\section{Quantisation of the dual action}
\renewcommand{\theequation}{3.\arabic{equation}}
\setcounter{equation}{0}

Let us now turn to a more detailed analysis of the dual action
(\ref{dualtheo}). To simplify notation, let us introduce the fields
\be \phi^\alpha = \left\{
\begin{array}{ll}
\psi^1 & \mbox{if $\alpha=1$,} \\
\t^\alpha &  \mbox{if $\alpha\neq 1$,}
\end{array} \right.
\ee
and denote 
$$\hat{d}_{\alpha \beta} = Tr (\hat{e}_\alpha \hat{e}_\beta) 
\hspace*{4cm} 
\hat{J}_\alpha = 2 i Tr (\d g g^{-1} \hat{e}_\alpha) \,.$$
The dual action is then, apart from the topological term which does
not matter for the following discussion,
\begin{eqnarray}
S^\natural[\phi^\alpha,g] & = & 
- \frac{k}{4 \pi} \int_{\M} dz d\bz\; Tr (g^{-1} \d g\; g^{-1} \bd g)
+ \frac{k}{12 \pi} \int_{\B} d^3 y \; \varepsilon^{ijk} \;
Tr(g^{-1} \d_i g \; g^{-1} \d_j g \; g^{-1} \d_k g) \nonumber \\
& & \;\;\;
+ \frac{k}{4 \pi} \int_{\M} dz d\bz \; \hat{d}_{\alpha \beta} \;
\d \phi^\alpha \bd \phi^\beta 
- \frac{k}{2\pi} \int_{\M} dz d\bz \; i \bd \phi^\alpha 
Tr (\d g g^{-1} \hat{e}_{\alpha})\,.
\end{eqnarray}
We now calculate the variation of the action with respect to a variation
of $\phi^\alpha$ and $g$, and find
\be
\delta S^\natural = \frac{k}{2\pi}
\int_{\M} dz d \bz \; Tr \left( \left[i \delta \phi^\alpha \hat{e}_\alpha 
+ e^{i \phi^\beta \hat{e}_\beta} \delta g 
g^{-1} e^{- i \phi^\beta \hat{e}_\beta} \right]
\bd \left( i \d \phi^\beta \hat{e}_\beta + e^{i \phi^\beta \hat{e}_\beta}
(\d g g^{-1}) e^{-i \phi^\beta \hat{e}_\beta} \right) \right)\,,
\ee
which implies the equation of motion 
\be
\label{eqmo}
\bd \J = 0 \,,
\ee
where
\be
\J= \left( i \d \phi^{\alpha} \hat{e}_\alpha 
+ e^{i \phi^\beta \hat{e}_\beta}
(\d g g^{-1}) e^{-i \phi^\beta \hat{e}_\beta} \right) \,.
\ee
If we also define
\be
\bJ= \left( g^{-1} i \bd \phi^\alpha \hat{e}_\alpha g 
+ g^{-1} \bd g \right) \,,
\ee
then
\be
\bd \J = e^{i \phi^\beta \hat{e}_\beta} g \; \d \bJ \; g^{-1}
e^{-i \phi^\beta \hat{e}_\beta} \,,
\ee
and thus $\d \bJ=0$ is equivalent to (\ref{eqmo}). It is then clear
that the two currents $\J(z)$ and $\bJ(\bz)$ only depend on one of the 
two variables. In terms of $g_0$ they are simply given as
\be
\label{g0ex}
\J = \d g_0 g_0^{-1} \hspace*{4cm}
\bJ = g_0^{-1} \bd g_0 \,,
\ee
where
$$ g_0 = e^{i \phi^{\alpha} \hat{e}_\alpha} g \,. $$

Following Witten \cite{Wit}, we want to calculate the Poisson brackets
of these currents. It is clear that the Poisson bracket of $\J$ with
$\bJ$ vanishes, and that the calculation for $\J$ and $\bJ$ is
essentially identical. Let us introduce
\be
\J^a (z)= Tr \left( t^a \J(z) \right) \hspace*{4cm}
\J^b (z')= Tr \left( t^b \J(z') \right) \,,
\ee
and likewise
\be
\bJ^a (\bz)= Tr \left( t^a \bJ(\bz) \right) \hspace*{4cm}
\bJ^b (\bz')= Tr \left( t^b \bJ(\bz') \right) \,.
\ee
We then find
\be
\delta \J^a = Tr \left( g^{-1} e^{-i \phi^\alpha \hat{e}_\alpha} t^a 
e^{i \phi^\alpha \hat{e}_\alpha}g \;
\d \left( g^{-1} i \delta \phi^{\alpha} \hat{e}_\alpha g
+ g^{-1} \delta g \right) \right) \,,
\ee
and derive, as in \cite{Wit}, the Poisson brackets 
\be
\left\{\J^a,\J^b\right\} = - \frac{2 \pi}{ k} \delta'(z-z') Tr (t^a t^b) 
+ \frac{2 \pi}{k} \delta(z-z') Tr \left( [t^a,t^b] \J(z) \right) \,,
\ee
and similarly for the bared currents. If we define now
\be
J^a(z) = - \frac{ik}{2\pi} 
Tr \left( t^a (i \d \phi^\alpha \hat{e}_\alpha
+ e^{i \phi^\beta \hat{e}_\beta} (\d g g^{-1})
e^{-i \phi^\beta \hat{e}_\beta})\right) = - \frac{ik}{2\pi} \J^a(z) \,,
\ee
and
\be
\bar{J}^a(\bz) = \frac{ik}{2\pi} 
Tr \left( t^a (g^{-1} i \bd \phi^\alpha \hat{e}_\alpha g
+ g^{-1} \bd g )\right) = \frac{ik}{2\pi} \bJ^a(\bz) \,,
\ee
then these currents satisfy the Poisson bracket relations,
\be
\left\{ J^a(z),J^b(z')\right\} 
= f^{abc} J^c(z) \delta(z-z') +
\frac{k}{4 \pi} \delta^{ab} \delta'(z-z') \,,
\ee
and
\be
\left\{ \bar{J}^a(\bz),\bar{J}^b(\bz')\right\} 
= f^{abc} \bar{J}^c(\bz) \delta(\bz-\bz') +
\frac{k}{4 \pi} \delta^{ab} \delta'(\bz-\bz') \,,
\ee
where $[t^a,t^b]=if^{abc} t^c$ and $Tr(t^a t^b)=\half \delta^{ab}$.

Next we want to expand the currents in modes, and in order to be able
to do so, we have to analyse their monodromy. We can use the fact that
the currents can be written in terms of $g_0$ as in (\ref{g0ex}), and
that the monodromy of $g_0$ is $g_0 \mapsto h_p g_0$, where
\be
\label{twists}
h_p = \exp\left(2 \pi i \eta_p e^1\right) \hspace*{1cm} \mbox{with} 
\hspace*{1cm} \eta_p = \frac{p}{k} \,,
\ee
and $p$ is an integer. It is then easy to see that the monodromy of
$\bar{J}^a(\bz)$ is trivial for each $p$, as 
\be
\bJ(\bz) \mapsto g_0^{-1} h_p^{-1} \bd h_p \,g_0 = g_0^{-1} \bd g_0 = 
\bJ(\bz) \,,
\ee
and we can hence expand $\bar{J}^a(\bz)$ in modes
\be
\bar{J}^a(\bz) = \sum_{n\in\Zop} \bar{J}^a_n \bar{\z}^{-n-1} \,,
\ee
where $\bar{\z} = \exp(2 \pi i \bz)$. Upon quantising the Poisson
brackets, the modes then satisfy the (untwisted) Kac-Moody (KM) algebra
\be
\label{KMcomm}
{} [\bar{J}^a_m,\bar{J}^b_n] = i f^{abc} \bar{J}^c_{m+n} 
+ \half \, k \, m \, \delta^{ab} \delta_{m,-n} \,.
\ee

For the following it is useful to write this algebra in a different
basis, the modified Cartan-Weyl basis \cite{GO}. In this basis the
algebra is generated by $\bH^i_m$, $i=1,\ldots r$, $m\in\Zop$, where
$r$ is the rank of $\lg$, and $\bE^\alpha_n$, $n\in\Zop$, where 
$\alpha$ parametrises the roots. The commutation relations are 
given as (see for example \cite{GO}) 
\be
\label{KMcomm1}
\begin{array}{ccl}
{\displaystyle [\bH^i_m, \bH^j_n]} & = & 
{\displaystyle k \, m \, \delta^{ij} \delta_{m,-n}} \vspace*{0.3cm}   \\
{\displaystyle [\bH^i_m, \bE^{\alpha}_n]} & =  & 
{\displaystyle \alpha^i \bE^{\alpha}_{m+n}} \vspace*{0.3cm}   \\
{\displaystyle [\bE^{\alpha}_m, \bE^{\beta}_n]} & = &
\left\{
\begin{array}{ll}
{\displaystyle \varepsilon(\alpha,\beta) \bE^{\alpha+\beta}_{m+n}}
\vspace*{0.2cm} & 
\mbox{if $\alpha+\beta$ is a root,} \vspace*{0.1cm} \\
{\displaystyle \frac{2}{\alpha^2} \left( \alpha \cdot \bH_{m+n} 
+ k \, m \, \delta_{m,-n} \right)} & 
\mbox{if $\alpha=-\beta$,} \vspace*{0.2cm}  \\
{\displaystyle 0} & \mbox{otherwise.}   
\end{array} \right. \vspace*{0.4cm}  \\
{}[k,\bE^{\alpha}_n] & =  & [k,\bH^i_n] = 0\,.
\end{array}
\ee

Let us now turn to the mode expansion of the other current. Using the
same reasoning as above, we see that the monodromy of $\J(z)$ is given
by 
\be
\label{mono}
\J(z) \mapsto \d (h_p g) g^{-1} h_p^{-1} = h_p \, \J(z) \, h_p^{-1} \,.  
\ee
If we write the various components of $J^a(z)$ in the modified
Cartan-Weyl basis of the Lie algebra (corresponding to the above basis
of the KM algebra), the adjoint action of $h_p$ is diagonal.
For each sector described by the monodromy $h_p$, we can then
expand the currents
\be
J^i(z) = \sum_{n\in\Zop} H^i_n \bar{\z}^{-n-1} \,, \hspace*{3cm}
J^{\alpha}(z) = \sum_{m\in\Zop + \langle\chi_p, \alpha\rangle} 
E^{\alpha}_m \bar{\z}^{-m-1}
\ee
where $\z = \exp(2 \pi i z)$, $\chi_p = \eta_p e_1$, and, to follow
more conventional notation, $\langle \cdot , \cdot \rangle$ denotes the
Killing form. Upon quantisation, these modes satisfy formally the same
commutation relations as (\ref{KMcomm1}), but the indices $m$ are not
all integral. This algebra is called the twisted KM algebra; the twist
is inner and is given by (\ref{twists}).
\medskip

We thus find that the dual model has various sectors which are
parametrised by the monodromy of the field $\phi^1$. In each sector,
the theory has two commuting infinite-dimensional symmetry algebras:
the untwisted (right-moving) KM algebra $\bar{J}^a_n$ and the
twisted (left-moving) KM algebra $H^i_n, E^\alpha_m$. 

It is then clear, that the space of states is of the form
\be
\H^\natural = \oplus_{p} \H^\natural_p \,,
\ee
where $p$ denotes the different sectors, and each $\H^\natural_p$
is a subspace of
\be
\H^\natural_p  \subset \H^{(p)} \otimes \H \,,
\ee
where $\H$ and $\H^{(p)}$ are the direct sums of all (irreducible)
representations of the untwisted and the twisted KM algebra (with the
twist corresponding to $h_p$), respectively. In general, we expect
there to be different quantisations, which correspond to different
(consistent) choices for the spaces $\H^\natural_p$, {\it cf.}\
\cite{MG5}. However, as we shall show in the next section, there
exists one quantisation for which the spectrum of the dual theory
coincides with the original one, and this suggests that this theory
is indeed equivalent to the original theory.

\section{Representations of the twisted Kac-Moody algebra}
\renewcommand{\theequation}{4.\arabic{equation}}
\setcounter{equation}{0}

We now want to analyse the representation theory of the twisted
KM algebra. The crucial property which will allow us to do this
rather easily is the fact that the above twisted KM algebra
is isomorphic to the untwisted algebra. This isomorphism is
explicitly known, and can be defined as
\begin{eqnarray}
\varphi(F^\alpha_m) & = & 
E^\alpha_{m+ \langle \chi_p,\alpha\rangle} \nonumber \\
\label{isom}
\varphi(I^i_n) & = & H^i_n + k \, \chi_p^i \delta_{n,0} \,,
\end{eqnarray}
where $H^i_n, E^\alpha_m$ are the generators of the twisted KM algebra
corresponding to the $p$-th sector, and $I^i_n, F^\alpha_m$ are
the generators of the untwisted KM algebra \cite{GO}. It is also
clear, that the inverse exists, and that it is given by 
\begin{eqnarray}
\varphi^{-1}(E^\alpha_m) & = & 
F^\alpha_{m - \langle \chi_p,\alpha\rangle} \nonumber \\
\varphi^{-1}(H^i_n) & = & I^i_n - k \, \chi_p^i \delta_{n,0} \,.
\end{eqnarray}

In particular, this implies that the representations of the twisted
and the untwisted KM algebra are in one-to-one correspondence. We
shall therefore denote the states in the corresponding representation
spaces by the same symbol.

The untwisted KM algebra contains a Virasoro algebra 
\be
{}[L_m,L_n]=(m-n) L_{m+n} + \frac{c}{12} m (m^2 -1) \delta_{m+n,0} 
\ee
in its universal enveloping algebra by virtue of the
Sugawara construction, where the generators of
the Virasoro algebra are given as
\be
L_n = \frac{1}{2 k+Q_\psi} \sum_{m\in\Zop}
\left( \sum_{i=1}^{r} : I^i_n I^i_{n-m} : + 
\sum_{\alpha} \frac{\alpha^2}{2} : F^{\alpha}_n F^{-\alpha}_{n-m}:
\right)\,.
\ee
Here $Q_\psi$ is the quadratic Casimir in the adjoint representation,
$: \cdot :$ denotes the usual normal ordering (see {\it e.g.}\ \cite{GO}),
and 
\be
c= \frac{k \dim \lg}{2k + Q_\psi} \,.
\ee
These generators are the modes of the stress-energy tensor, as
\be
{}[L_m,I^i_n] = -n I^i_{m+n} \hspace*{3cm} [L_m,F^{\alpha}_n] 
= -n F^{\alpha}_{m+n} \,.
\ee
Using the above isomorphism, it is then clear that the $\varphi(L_n)$
satisfy a Virasoro algebra with the same central charge, but
that they are not quite the mode expansion of the stress-energy tensor
as
\begin{eqnarray}
{}[\varphi(L_m),E^\alpha_n] & = &
- (n - \langle \chi_p,\alpha\rangle) E^\alpha_{m+n} \nonumber \\
{}[\varphi(L_m),H^i_n] & = &
- n H^i_{m+n} - k \, n \, \chi_p^i \delta_{m+n,0} \nonumber \,.
\end{eqnarray}

However, it is easy to see how to modify this by defining ({\it c.f.}\
for example \cite{FH})
\begin{eqnarray}
L_m^{(p)} & = &
\varphi(L_m) - \chi_p^i H^i_m - \half k\, \chi_p^i \chi_p^i \delta_{m,0} 
\nonumber \\
& = &
\varphi 
\left( L_m - \chi_p^i I^i_m + \half k \, \chi_p^i \chi_p^i \delta_{m,0} 
\right) \,.
\end{eqnarray}
Then the $L_m^{(p)}$ define a Virasoro algebra with the same central charge
as before, and
\be
{}[L_m^{(p)},E^\alpha_n]= -n E^\alpha_{m+n} \hspace*{3cm}
{}[L_m^{(p)},H^i_n] = - n H^i_{m+n} \,.
\ee
In particular, this implies that the dual theory is a conformal field
theory, as the theory contains a conformal stress-energy tensor.
\medskip

We now want to analyse whether there exists a subspace $\H^\natural_p$ for
each sector (labelled by $p$), so that the spectrum of the dual theory
is the same as the original one. We observe that if we consider
in the $p$-th sector only states which satisfy   
\be
\label{secp}
\frac{2}{\langle e_1,e_1 \rangle} \; \left( e_1 \cdot H_0 \right) \; \mu 
= - p \; \mu \,,
\ee
then
\be
\frac{2}{\langle e_1,e_1 \rangle} \; \left( e_1 \cdot I_0 \right) \; \mu 
= p  \; \mu \,,
\ee
and
\be
L_0^{(p)} \mu  = L_0 \;  \mu  \,,
\ee
as 
\begin{eqnarray}
\left( - \left( \chi_p \cdot I_0 \right)
+ \half \; k \; \chi_p^2 \right)  \mu & = &
\frac{p}{k} \left( - \left( e_1 \cdot I_0 \right)
+ \frac{\langle e_1,e_1 \rangle}{2} p \right)  \mu \nonumber \\
& = & 
\frac{p \; \langle e_1, e_1 \rangle}{2 \, k}
\left( - p + p \right) \mu  = 0 \,.
\end{eqnarray}

This is not yet a complete solution, as we should restrict $p$ to lie
in $0,1,\ldots, \frac{2 k} {\langle e_1,e_1\rangle} - 1$, since the
monodromy $h_p$ corresponding to $p$ and $p+\frac{2 k}{\langle
e_1,e_1\rangle}$ is the same. (The coroot $2 e_1 / \langle e_1,
e_1\rangle$ is mapped to the unit element of the group under the
exponential map.) On the other hand, in the spirit of orbifold
constructions \cite{DHVW}, we should expect that the states in each
sector are characterised by their transformation property under the
group of monodromies $\{h_p \}$. We should therefore consider in the
$p$-th sector all states which satisfy
\be
\label{secp1}
\frac{2}{\langle e_1,e_1 \rangle}\;  \left( e_1 \cdot H_0 \right) \;  \mu 
= - \left(p + \frac{2 \, k \, m}{\langle e_1,e_1 \rangle} \right)
\mu \,,
\ee
where $m\in\Zop$. Then we find that 
\be
\label{muI}
\frac{2}{\langle e_1,e_1 \rangle}\;  \left( e_1 \cdot I_0 \right) \; \mu 
= \left(p - \frac{2 \, k \, m}{\langle e_1,e_1 \rangle} \right)
\mu  \,,
\ee
and
\be
\label{muL}
L_0^{(p)} \mu = L_0 \; \mu + p \, m \, \mu \,.
\ee

To see that the spectrum of $L_0$ is really the same, let us now
recall that the affine Weyl group contains the subgroup of coroot
translations, which act on the generators as \cite{GO}
\begin{eqnarray}
I_m^i & \mapsto & I_m^i(c):= I_m^i - k c^i \delta_{m,0} \nonumber 
\vspace*{0.2cm} \\
F^\alpha_n & \mapsto & 
F^\alpha_n(c):= F^\alpha_{n - \langle \alpha, c \rangle} 
\nonumber \\
L_m & \mapsto & L_m(c):= L_m  - c^i I^i_m + \half k \, c^i c^i 
\delta_{m,0} \,,
\end{eqnarray}
where $c$ is a coroot. If we apply this transformation with
\be
c = - \frac{2}{\langle e_1, e_1 \rangle} m e_1
\ee
to the above state, satisfying (\ref{muI}), we find that 
\be
\frac{2}{\langle e_1,e_1 \rangle} \; \left(e_1 \cdot I_0(c)\right)
\; \mu = \left( p + \frac{2 k m}{\langle e_1,e_1 \rangle} 
\right) \mu \,,
\ee
and
\begin{eqnarray}
L_0(c) \; \mu & =  &
\left( L_0 + \frac{2}{\langle e_1,e_1\rangle } m \; 
\left(e_1 \cdot I_0\right)
+ \half k \frac{4}{\langle e_1,e_1\rangle } m^2 \right) \mu 
\nonumber \\
& = & 
\left (L_0 + m 
\left( p - \frac{2 k m}{\langle e_1,e_1 \rangle} \right)
+ k m^2 \frac{2}{\langle e_1,e_1\rangle } \right) \mu 
\nonumber \\
& = & L_0 \; \mu + p \,m \;\mu \,.
\end{eqnarray}
This is precisely the right-hand-side of (\ref{muL}). 
We can thus define $\H^\natural_p$ to be the span
\be
\H^\natural_p = \left\langle (\mu \otimes \nu) \in\H_i\otimes\H_i
\left| \frac{2}{\langle e_1, e_1 \rangle}\; \left(e_1 \cdot I_0
\right)\, \mu =  \left(p +  \frac{2 k m}{\langle e_1,e_1
\rangle} \right) \mu; \; m\in\Zop, i\in R(\hat{\lg}) \right. 
\right\rangle \,,
\ee
where $R(\hat{\lg})$ is the set of irreducible highest weight
representations of the KM algebra $\hat{\lg}$, and $\mu$ is to be
regarded as an element of the twisted representation space by virtue
of the isomorphism (\ref{isom}). This gives rise to the consistent
theory 
\be
\H^{dual} = \bigoplus_{p=0}^{\frac{2 k}{\langle e_1,e_1\rangle} - 1}
\H_p^\natural \,,
\ee
as the definition respects the multiplicative structure of the field
theory, {\it i.e.}\  the fusion product of states in
$\H^\natural_{p_1}$ and $\H^\natural_{p_2}$ lies in $\H^\natural_{p}$,
where $p=p_1 + p_2 \;\;\mbox{mod}\;\; 2 k /\langle e_1,e_1\rangle$. It
is clear by construction, that this dual theory  has the same spectrum
as the original theory.

\section{Conclusions}
\renewcommand{\theequation}{5.\arabic{equation}}
\setcounter{equation}{0}

We have analysed the (left-)chiral abelian duality transformation of a WZW
model corresponding to a simple, simply-connected group. We have shown
that the dual theory possesses two infinite dimensional symmetry
algebras, a (right-moving) untwisted, and a (left-moving) twisted KM
algebra. We have analysed the spectrum of this theory, and have shown
that there exists a quantisation for which the spectrum of the
original theory is recovered. This dual theory can naturally be
interpreted as an orbifold \cite{DHVW} of the original model induced
by the twists (\ref{twists}).
\smallskip

It would be interesting to do a similar analysis for various, more
general cases: (i) the case, where the field $\t^1$ with respect to
which the transformation is performed is not purely left (or right)
moving, and (ii) the case, where the fields $\t^i$ span an arbitrary,
not necessarily abelian subgroup of the target space group. (This
latter case is usually referred to as non-abelian duality
\cite{AAL1}.) 

It would also be interesting to understand which modifications arise
if the target space group is not simply-connected. 
\bigskip

\noindent {\bf Acknowledgements}

It is a pleasure to thank Peter Goddard for many enlightening
discussions. I also thank John Varghese for useful conversations at an
early stage of this work, George Papadopoulos for helpful discussions,
and G\'erard Watts for a useful remark.

This work was supported by a Research Fellowship of Jesus College,
Cambridge, and partly by PPARC and EPSRC, grant GR/J73322.


\begin{thebibliography}{[20]}

\bibitem {AABL} Alvarez E., Alvarez-Gaum\'e L., Barb\'on J.L.F.,
Lozano Y.: Some global aspects of duality in string
theory. Nucl. Phys.  {\bf B~415}, 71-100 (1994)

\bibitem {AAL1} Alvarez E., Alvarez-Gaum\'e L., Lozano Y.:
On non-abelian duality. Nucl. Phys. {\bf B~424}, 155-183 (1994)

\bibitem {AAL2} Alvarez E., Alvarez-Gaum\'e L., Lozano Y.: A
canonical approach to duality transformations. Phys. Lett.  
{\bf B~326}, 183-189 (1994)

\bibitem {AAL3} Alvarez E., Alvarez-Gaum\'e L., Lozano Y.: An
introduction to $T$-duality in string theory. preprint
CERN-TH-7486-94, hep-th/9410237

\bibitem {Buscher} Buscher T.: A symmetry of the string background
field equations. Phys. Lett. {\bf B~194}, 59-62 (1987); Path-integral
derivation of quantum duality in nonlinear
sigma-models. Phys. Lett. {\bf B~201}, 466-472 (1988)  

\bibitem{MPHOS} Chu M., Goddard P., Halliday I., Olive D.,
Schwimmer A.: Quantisation of the Wess-Zumino-Witten
model on a circle. Phys. Lett. {\bf B~266}, 71-81 (1991); 
Papadopoulos G., Spence B.: The canonical structure of
Wess-Zumino-Witten models. Phys. Lett. {\bf B~292}, 321-328 (1992)

\bibitem{DHVW} Dixon L., Harvey J., Vafa C., Witten E.: Strings on
orbifolds. Nucl. Phys. {\bf B 261}, 678-686 (1985); Strings on
orbifolds (II). Nucl. Phys. {\bf B 274}, 285-314 (1986)

\bibitem{FGK} Felder G., Gawedzki K., Kupiainen A.: Spectra of
Wess-Zumino-Witten models with arbitrary simple groups.  Commun. Math.
Phys. {\bf 117}, 127-158 (1988)

\bibitem{FH} Freericks J.K., Halpern, M.B.: Conformal deformation by
the currents of affine $g$. Ann. Phys.~{\bf 188}, 258-306 (1988) 

\bibitem{FKS} Frenkel I.B., Kac V.G.: Basic representations of affine
Lie algebras and dual resonance models. Invent. Math. {\bf 62}, 23-66 
(1980); Segal G.: Unitary representations of some infinite dimensional
groups. Commun. Math. Phys. {\bf 80}, 301-342 (1981)

\bibitem{MG5} Gaberdiel M.R.: WZW models of general simple groups.
hep-th/9508105, DAMTP-95-46, to appear in Nucl. Phys. {\bf B}

\bibitem{Ginsparg} Ginsparg P.: Curiosities at $c=1$. Nucl. Phys. 
{\bf B~295}, 153-170 (1988)

\bibitem{GNOS} Goddard P., Nahm W., Olive D., Schwimmer A.: Vertex
operators for non-simply-laced algebras. Commun. Math. Phys. 
{\bf 107}, 179-212 (1986)

\bibitem{GO} Goddard P., Olive D.: Kac-Moody and Virasoro algebras in
relation to quantum physics. Int. Journ. Mod. Phys. {\bf A~1}, 303-414
(1986) 

\bibitem{Kir} Kiritsis E.: Exact duality symmetries in CFT and string
theory. Nucl. Phys. {\bf B~405}, 109-142 (1993)

\bibitem{PW} Polyakov A.M., Wiegmann P.B.: Goldstone fields in two
dimensions with multivalued actions. Phys. Lett. {\bf B~141}, 
223-228 (1984) 

\bibitem{RV} Ro\v{c}ek M., Verlinde E.: Duality, quotients, and
currents. Nucl. Phys. {\bf B~373}, 630-646 (1992)

\bibitem{S} Seiberg N.: Electric-Magnetic duality in supersymmetric
non-abelian gauge theories. Nucl. Phys. {\bf B~435}, 129-146 (1995)

\bibitem{SW} Seiberg N., Witten E.: Electric-magnetic duality,
monopole condensation, and confinement in N=2 supersymmetric
Yang-Mills theory. Nucl. Phys. {\bf B~426}, 19-52 (1994)

\bibitem{Wit} Witten E.: Non-abelian bosonization in two dimensions.
Commun. Math. Phys. {\bf 92}, 455-472 (1984)

\bibitem{W2} Witten E.: String theory dynamics in various dimensions.
Nucl. Phys. {\bf B~443}, 85-126 (1995)


\end{thebibliography}
\end{document}